\documentclass[12pt]{article}
\usepackage{amsmath}
\usepackage{amssymb}
\usepackage[mathscr]{eucal}
\usepackage[usenames]{color}
\usepackage[latin5]{inputenc}
\usepackage{url}

\newtheorem{thm}{Theorem}

\numberwithin{equation}{section}
\numberwithin{thm}{section}
\numberwithin{lemma}{section}
\numberwithin{prop}{section}
\numberwithin{cor}{section}
\numberwithin{rmk}{section}
\numberwithin{defn}{section}

\definecolor{darkolivegreen}{rgb}{0.333333, 0.419608, 0.1843140}

\newcommand{\dx}{\partial_x}
\newcommand{\dy}{\partial_y}
\newcommand{\dz}{\partial_z}
\newcommand{\dt}{\partial_t}
\newcommand{\du}{\partial_u}
\newcommand{\dv}{\partial_v}
\newcommand{\dw}{\partial_w}

\begin{document}

\title{\Large Davey-Stewartson Equations in (3+1)-Dimensions with an \\ Infinite Dimensional Symmetry Algebra
}

\author{
C. \"{O}zemir\\
\small Department of Mathematics, Faculty of Science and Letters,\\
\small Istanbul Technical University, 34469 Istanbul,
Turkey\thanks{e-mail: ozemir@itu.edu.tr} }

\date{24 February 2019}

\maketitle

\begin{abstract}
This article is devoted to discovering  Lie symmetry algebra of a (3+1)-dimensional Davey-Stewartson system which appears in the field of plasma physics. It is found that the algebra is an infinite dimensional one and of Kac-Moody type. Making use of these symmetries, some reduced equations to lower dimensions are also presented.
\end{abstract}

\section{Introduction}
The aim of this paper is to investigate the Lie symmetry algebra of the  Davey-Stewartson (DS) equations in (3+1)-dimensions which we consider in the form
\begin{subequations}\label{3DDS}
\begin{eqnarray}
   \label{DDSa}  &&i\psi_t+\psi_{xx}+a_1\psi_{yy}+\psi_{zz}=a_2|\psi|^2\psi+\psi w, \\
   \label{DDSb}  &&w_{xx}+b_1w_{yy}+w_{zz}=b_2(|\psi|^2)_{yy}
\end{eqnarray}
\end{subequations}
where all of the coefficients are assumed to be nonzero real constants. $\psi$ is a complex and $w$ is a real function of $t,x,y,z$. This equation appears in \cite{carbonaro2012three} in the form
\begin{subequations}\label{carbo}
\begin{eqnarray}
     &&i\Phi_{\tau}+\Phi_{xx}+\Phi_{yy}-\frac{3\omega^2}{\beta}\Phi_{zz}=-q|\phi|^2\Phi-\Phi U, \\
     &&U_{xx}+U_{yy}+(1-\frac{\lambda^2}{\beta})U_{zz}=2\nu^2(|\Phi|^2)_{zz}
\end{eqnarray}
\end{subequations}
through  three-dimensional modulation of an electron-acoustic wave in a collisionless unmagnetized plazma via multiple scales asymptotic expansion method. According to their analysis, the far-field of the propagation of a
monochromatic modulation wave in this plasma consisting of  two species of electrons at different
temperatures is governed by this Davey-Stewartson type system \cite{carbonaro2012three}.  The transition between  \eqref{3DDS} and \eqref{carbo} is easy with the replacement $w\leftrightarrow -U$ and $y\leftrightarrow z$.

Davey-Stewartson system  was first obtained in \cite{daveystewartson} in two space dimensions by a perturbation expansion as governing equations for the evolution of a wave packet in a water of finite depth. It also appears as a special case of the Benney-Roskes system \cite{benneyroskes}. Integrability of the Davey-Stewartson system through inverse scattering transform was studied in \cite{fokas1983inverse}. \cite{fokas1984inverse} is to be mentioned in the same sense. Not restricted to the integrable case, the well-posedness of the initial value problem was studied in \cite{saut1990initial} and was followed by \cite{linares1993davey}. We would also like to note that \cite{saut1990initial} includes expository lines on the early literature on DS equations.

Considering Lie symmetry algebras, \cite{champagnewinternitz1988} shows that the DS system in (2+1)-dimensions
\begin{subequations}\label{2DDS}
\begin{eqnarray}
   \label{2DDSa}  &&i\psi_t+\psi_{xx}+\epsilon_1\psi_{yy}=\epsilon_2|\psi|^2\psi+\psi w, \\
   \label{2DDSb}  &&w_{xx}+\delta_1w_{yy}=\delta_2(|\psi|^2)_{yy}
\end{eqnarray}
\end{subequations}
has an infinite-dimensional Lie symmetry algebra of Kac-Moody-Virasoro (KMV) type exactly in the integrable case; i.e., when $\delta_1=-\epsilon_1=\mp 1$. The symmetry algebra is generated by four vector fields of first order differential operators which include four arbitrary functions of time. This is typical (albeit not necessary) among the well known integrable equations in (2+1)-dimensions. Detailed literature and results in this direction can be found in the Refs. \cite{gungor2001symmetries,gungor2002generalized,gungor2004equivalence,gungor2006generalized,gungor2006virasoro,gungor2010infinite,basarab2013infinite,
gungor2016variable}. In addition to the literature focusing on certain equations or family of equations, the inverse problem of determining classes of equations in (3+1)-dimensions that have infinite-dimensional symmetry algebras of Virasoro type is studied in \cite{lin2001high}. As the inverse problem again, the realizations of the Witt and Virasoro  algebras by Lie vector fields of first order differential operators in $\mathbb{R}^n$, $n\leq 3$ is put into stage in \cite{zhdanov2014realizations}, presenting in a sophisticated, systematic way \emph{the (1+1)-dimensional nonlinear PDEs which are integrable in the sense that they admit infinite symmetries of Virasoro type} \cite{zhdanov2014realizations}.

Regarding the literature on the derivation of DS equations, a (2+1)-dimensional  system in the context of plasma physics was obtained in \cite{nishinari1993} by  a perturbation approach. The evolution equations   derived in \cite{nishinari1994multi}, numbered (17,18), to describe wave-packet dynamics of an ion wave in a magnetic field are in (3+1)-dimensions and reduce to a planar DS-like system when  the third dimension is ignored. In \cite{annou2012dromion}, we see a similar system of equations in (3+1)-dimensions and one of the space dimensions is dropped from the analysis afterwards. Another (2+1)-dimensional model appears in \cite{nishinari1994new}. The nonlinear modulation of an electromagnetic localized pulse yields a three-dimensional generalization of the Davey-Stewartson system in \cite{leblond1999electromagnetic}.  By a perturbative reduction method, starting with a known integrable equation in (2+1)-dimensions, a generalization of the DS equations to four field variables is obtained in \cite{maccari2001class}. Refs. \cite{yuan2009symmetry} and \cite{xiao2011full} are devoted to the determination of symmetries and similarity solutions of the resonant DS system, which is a generalization of the resonant nonlinear Schr\"{o}dinger equation. The global existence and blow up for (3+1)-dim. DS equations is studied in \cite{li2016global}. Let us also mention the 2-d case model \cite{panguetna2017two} and 3-d derivation \cite{tabi2018electronegative}.

To our knowledge, there is no result in the literature on the symmetry algebra of the DS equations in (3+1)-dimensional case. Motivated by the physical literature above, we aim at studying the symmetries in this case. The system \eqref{3DDS}, the same as  \eqref{carbo} put in the literature by \cite{carbonaro2012three}, we choose as our prototype equation to work on, as it gives an obvious possibility to compare the results we obtain with the results on \eqref{2DDS} in \cite{champagnewinternitz1988} in the (2+1)-case.

Therefore, this article is based on answering the following questions:
\begin{itemize}
\item What is the Lie symmetry algebra of the system \eqref{3DDS}? Is the KMV symmetry algebra available in the absence of the $z$-variable conserved or is it broken?
\item What are possible reductions  of \eqref{3DDS}  to equations with lower number of independent  variables and is any of them integrable?
\end{itemize}
In Section 2 we answer the first question and Section 3 is devoted to the latter one.

\section{Lie Point Symmetries}\label{section3}
Writing $\psi=u+iv$ we express \eqref{3DDS} as the following system
\begin{subequations}\label{Triple}
\begin{eqnarray}
   \label{triple1}  &&\Delta_1=u_t+v_{xx}+a_1v_{yy}+v_{zz}-a_2v(u^2+v^2)-vw=0, \\
   \label{triple2}  &&\Delta_2=-v_t+u_{xx}+a_1u_{yy}+u_{zz}-a_2u(u^2+v^2)-uw=0,\\
   \label{triple3}  &&\Delta_3=w_{xx}+b_1w_{yy}+w_{zz}-b_2(u^2+v^2)_{yy}=0.
\end{eqnarray}
\end{subequations}
The infinitesimal generator of the Lie symmetry algebra of the system is of the form
\begin{equation}
V=\tau\dt+\xi\dx+\eta\dy+\zeta \dz+\phi_1\du+\phi_2\dv+\phi_3\dw.
\end{equation}
The vector field  $V$
with coefficients depending on $(t,x,y,u,v,w)$  is a symmetry generator if its second order prolongation is an annilihator of \eqref{Triple} on its solution surface:
\begin{equation}\label{inf-inv}
  \mathsf {Pr}^{(2)}V(\Delta_i)\big\vert_{\Delta_j=0}=0,\quad i,j=1,2,3.
\end{equation}
Skipping the intermediate calculations, we find that, $c_1,c_2,c_3$ being arbitrary constants,
\begin{equation}
V=c_1X_1+c_2X_2+c_3X_3+Y(g)+Z(h)+Q(k)+W(m),
\end{equation}
with
\begin{subequations}
\begin{eqnarray}
X_1&=&\dt,\\
X_2&=&2t\dt+x\dx+y\dy+z\dz-u\du-v\dv-2w\dw,\\
X_3&=&z\dx-x\dz,\\
Y(g)&=&g(t) \dx-\frac{x}{2}\big[\,g'(t)(v\du-u\dv)+g''(t)\dw\big]\\
Z(h)&=&h(t) \dy-\frac{y}{2a_1}\big[\,h'(t)(v\du-u\dv)+h''(t)\dw\big]\\
Q(k)&=&k(t) \dz-\frac{z}{2}\big[\,k'(t)(v\du-u\dv)+k''(t)\dw\big]\\
W(m)&=&m(t)(v\du-u\dv)+m'(t) \,\dw.
\end{eqnarray}
\end{subequations}
where $g(t)$, $h(t)$, $k(t)$ and $m(t)$ are arbitrary functions. The commutation relations are given below. The ones involving $X_i$'s, $i=1,2,3$,
\begin{align}
&[X_1,X_2]=2X_1, \quad [X_1,Y(g)]=Y(g'), \quad [X_1,Z(h)]=Z(h'), \nonumber\\
&[X_1,Q(k)]=Q(k'), \quad [X_1,W(m)]=W(m'),\nonumber\\
&[X_2,Y(g)]=Y(2tg'-g), \quad [X_2,Z(h)]=Z(2th'-h),\\
&[X_2,Q(k)]=Q(2tk'-k), \quad  [X_2,W(m)]=W(2tm'),\nonumber\\
&[X_3,Y(g)]=Q(g), \quad [X_3,Q(k)]=-Y(k), \nonumber
\end{align}
the remaining nozero relations,
\begin{align}
&[Y(g_1),Y(g_2)]=W\big(\frac{1}{2}(g_1'g_2-g_1g_2')\big),\nonumber\\
&[Z(h_1),Z(h_2)]=W\big(\frac{1}{2a_1}(h_1'h_2-h_1h_2')\big),\\
&[Q(k_1),Q(k_2)]=W\big(\frac{1}{2}(k_1'k_2-k_1k_2')\big), \nonumber
\end{align}
and finally, just to state explicitly, the vanishing commutations,
\begin{align}
&[X_1,X_3]=[X_2,X_3]=[X_3,Z(h)]=[X_3,W(m)]=0, \nonumber\\
&[Y(g),Z(h)]=[Y(g),Q(k)]=[Y(g),W(m)]=0,\\
&[Z(h),Q(k)]=[Z(h),W(m)]=[Q(k),W(m)]=[W(m_1),W(m_2)]=0. \nonumber
\end{align}
Having obtained these results, we can state the following.
\begin{thm}\label{thm1}
The symmetry algebra of the (3+1)-dimensional Davey-Stewartson system \eqref{3DDS} is infinite dimensional, with the structure
\begin{equation}
L=\{\{X_1,X_2\}\oplus X_3\}\rhd\{Y(g),Z(h),Q(k),W(m)\}
\end{equation}
where $A_{3.2}=\{X_1,X_2\}\oplus X_3$ is a three-dimensional Lie algebra and $\{Y(g),Z(h),Q(k),W(m)\}$ is a Kac-Moody algebra being a non-Abelian ideal.
\end{thm}
(For a list of low-dimensional Lie algebras besides $A_{3.2}$ see \cite{basarab2001structure}.) The system \eqref{3DDS} reduces to \eqref{2DDS} when the $z$-variable is ignored. According to the results available in \cite{champagnewinternitz1988}, \eqref{2DDS} has an infinite-dimensional symmetry algebra in the form $\hat{L}=\{X(f)\}\rhd\{Y(g),Z(h),W(m)\}$, with $Y,Z,W$ given above, where $f,g,h,m$ are arbitrary functions of time. Comparing with this result, we see that the generators $Y,Z,W$ are preserved in the (3+1)-dimensional case (we used the same lettering with \cite{champagnewinternitz1988} for the genarators). In addition, we have the generator $Q$, corresponding to the added variable $z$, and it is in the same fashion with $Y$ and $Z$. However, the Virasoro part $X(f(t))$ with arbitrary $f(t)$ no longer exists, it reduces to $\{X_1,X_2\}$, representing  translation and scaling, and there arises $X_3$, rotation around the $y$-axis (e.g. the rotation $x\leftrightarrow z$ is obvious in \eqref{3DDS}).

\section{Invariant Solutions}
In this section, we make use of the algebraic structure revealed above and  look for reductions of the system \eqref{3DDS} that determine solutions invariant under the group of transformations of a specific subalgebra.
\subsection{The finite dimensional subalgebra $\{X_1,X_2\}\oplus X_3$ }
The rotational symmetry of the equation around the $y$ axis is represented by the generator $X_3=z\dx-x\dz$. Any solution invariant under this symmetry will have the form
$\psi=\psi(r,y,t)$, $\quad w=w(r,y,t)$, where $r=\sqrt{x^2+z^2}$ which corresponds to looking for radial solutions in the $x$, $z$ coordinates and leaving the dependence on  other coordinates as they are.
The resulting system in  (1+1)-dimensions includes a radial term which can be eliminated by a transformation of the form $\psi(r,y,t)=\Psi(r,y,t)/\sqrt{r}$ , $w(r,y,t)=\hat{W}(r,y,t)/\sqrt{r}$.

The subalgebra $\{X_1,X_2\}\oplus X_3$ of the Lie algebra of the $3+1$-dimensional  system can be used to reduce it to a system of ODEs. To find the reduced equations we first pass to polar representation in the complex variable $\psi=u+iv$ as
$$\psi(x,y,z,t)=\rho(x,y,z,t)e^{i\phi(x,y,z,t)}$$
which converts \eqref{3DDS} to
\begin{subequations}\label{polar}
\begin{eqnarray}
   \label{}  -\rho\phi_t+\rho_{xx}+a_1\rho_{yy}+\rho_{zz}-\rho(\phi_x^2+a_1\phi_y^2+\phi_z^2)&=&a_2\rho^3+\rho w, \\
   \label{}  \rho_t+\rho(\phi_{xx}+a_1\phi_{yy}+\phi_{zz})+2\rho_x\phi_x+2a_1\rho_y\phi_y+2\rho_z\phi_z&=&0 \\
   \label{}  w_{xx}+b_1w_{yy}+w_{zz}&=&b_2(\rho^2)_{yy}.
\end{eqnarray}
\end{subequations}
We use the identities $u\du+v\dv=\rho \partial_{\rho}$ and $u\dv-v\du=\partial_{\phi}$ when necessary. Invariance under the generators $\{X_1,X_2,X_3\}$ yields that the dependent variables will have the form
\begin{equation}
\rho=\frac{1}{y}F(s), \quad \phi=G(s), \quad w= \frac{1}{y^2}\,M(s), \qquad s=\frac{x^2+z^2}{y^2}.
\end{equation}
The system \eqref{polar} reduces to
\begin{subequations}\label{red2}
\begin{eqnarray}
   \label{red2a}  4s(1+a_1s)(\ddot F-F\dot G^2) +2(2+5a_1 s) \dot F+2a_1 F&=&a_2F^3+MF\\
   \label{red2b}  2s(1+a_1s)F\ddot G+4s(1+a_1s)\dot F \dot G+(2+5a_1 s) F\dot G &=&0\\
                  s(1+b_1s)\ddot M+(1+\frac{7b_1}{2}s)\dot M +\frac{3b_1}{2}M&=&2b_2s^2(F\ddot F  + \dot{F}^2)\nonumber\\
   \label{red2c}  &+&7b_2sF\dot F+\frac{3b_2}{2}F^2
\end{eqnarray}
\end{subequations}
Eq. \eqref{red2b} integrates to
\begin{equation}
s(1+a_1s)^{3/2}F^2\dot G=C
\end{equation}
where $C$ is a constant.
\eqref{red2c} is integrated once as
\begin{equation}
s(1+b_1s)\dot M+\frac{3b_1}{2}sM=2b_2s^2F \dot F+\frac{3b_2}{2}sF^2+K_1,
\end{equation}
$K_1$ being an integration constant.
Let us set $F^2=N$, which amounts to saying $\displaystyle \rho=\frac{1}{y}\sqrt{N(s)}$. The system above is cast into
\begin{subequations}\label{olmadi}
\begin{eqnarray}
\label{olmadi1}M&=&s(1+a_1s)\Big[\,\frac{2\ddot{N}}{N}-\frac{\dot{N}^2}{N^2}\,\Big]-\frac{4C^2}{s(1+a_1s)^2}\frac{1}{N^2}+(2+5a_1s)\frac{\dot{N}}{N}+2a_1-a_2N,\\
\label{olmadi2}\dot M&+&b_1(s\dot M+\frac{3}{2}M)\,=\,b_2(s\dot N+\frac{3}{2}N)+\frac{K_1}{s}.
\end{eqnarray}
\end{subequations}
We could not find any way of integrating \eqref{olmadi2} to obtain $M$ in terms of $N$ to get a second order equation from \eqref{olmadi1}. When we plug $M$ in \eqref{olmadi2}, we get a third order equation in $N$ which is quite complicated to mention here and unfortunately we are without any step forward at this point. Nevertheless, the structure of the similarity variable $s$ resembles the form a lump solution variable and if one is able to find a solution of this system that may result in such a solution.

\subsection{Reduction through the infinite dimensional algebra}
Invariance under the infinite-dimensional symmetry generator
\begin{equation}\label{YZQW}
V=Y(g)+Z(h)+Q(k)+W(m)
\end{equation}
suggests that the dependent variables must have the form
\begin{align}
&\psi(x,y,z,t)=\exp \Big\{i\Big[\frac{g'}{4g}x^2+\frac{h'}{4a_1h}y^2+\frac{k'}{4k}z^2-\frac{m}{g}x\Big]\Big\}\Psi(\xi,\eta,t),\\
&w(x,y,z,t)=-\frac{g''}{4g}x^2-\frac{h''}{4a_1h}y^2-\frac{k''}{4k}z^2+\frac{m'}{g}x+\hat{W}(\xi,\eta,t)
\end{align}
where
\begin{equation}
\xi=h(t)x-g(t)y, \qquad \eta=k(t)x-g(t)z.
\end{equation}
Here we have assumed that the functions $g,h,k$ do not become zero. What we obtain is the reduction of \eqref{3DDS}
to (2+1)-dimensions
\begin{align}
i\Psi_t+(a_1g^2+h^2)\Psi_{\xi\xi}+2hk\Psi_{\xi\eta}+(g^2+k^2)\Psi_{\eta\eta}&\nonumber\\
+i\tilde{A}\Psi_\xi+i\tilde{B}\Psi_\eta+(i\tilde{C}+\tilde{D})\Psi&=a_2|\Psi|^2\Psi+\Psi \hat{W},\label{redgen1}\\
(b_1g^2+h^2)\hat{W}_{\xi\xi}+2hk\hat{W}_{\xi\eta}+(g^2+k^2)\hat{W}_{\eta\eta}&=b_2g^2(|\Psi|^2)_{\xi\xi}+\tilde{E}(t), \label{redgen2}
\end{align}
with
\begin{align}
\tilde{A}(\xi,t)&=(\frac{g'}{g}+\frac{h'}{h})\xi-2h\frac{m}{g},\nonumber\\
\tilde{B}(\eta,t)&=(\frac{g'}{g}+\frac{k'}{k})\eta-2k\frac{m}{g},\nonumber\\
\tilde{C}(t)&=\frac{g'}{2g}+\frac{h'}{2h}+\frac{k'}{2k},\\
\tilde{D}(t)&=-\frac{m^2}{g^2},\nonumber\\
\tilde{E}(t)&=\frac{g''}{2g}+\frac{b_1}{2a_1}\frac{h''}{h}+\frac{k''}{2k}. \nonumber
\end{align}
One possibility in the analysis of the system \eqref{redgen1}-\eqref{redgen2} is applying a coordinate transformation to eliminate the mixed derivative terms to check for any correspondence with the standard DS equations. However, such a transformation should be depending linearly both on $\xi$ and $\eta$ with time-varying coefficients, whereby producing extra differential terms on the right hand side of Eq. \eqref{redgen2} . Therefore, we could not set up any correspondence with the standard DS equations in this case (when $g,h,k$ are nonzero).

Now let us consider the special case $h(t)=0$ in \eqref{YZQW}. Invariance under the symmetry generator $V=Y(g)+Q(k)+W(m)$ gives the following reduction of \eqref{3DDS} to a (2+1)-dimensional system. The dependent variables will have the form
\begin{align}
&\psi(x,y,z,t)=\exp \Big\{i\big[\frac{g'}{4g}x^2+\frac{k'}{4k}z^2-\frac{m}{g}x\big]\Big\}\Psi(\zeta,y,t),\\
&w(x,y,z,t)=-\frac{g''}{4g}x^2-\frac{k''}{4k}z^2+\frac{m'}{g}x+\hat{W}(\zeta,y,t), \qquad \zeta=k(t)x-g(t)z,
\end{align}
after which we obtain
\begin{align}
i\Psi_t+A\Psi_{\zeta\zeta}+a_1\Psi_{yy}+iB\Psi_\zeta+(iC+D)\Psi&=a_2|\Psi|^2\Psi+\Psi \hat{W},\\
A\hat{W}_{\zeta\zeta}+b_1\hat{W}_{yy}&=b_2(|\Psi|^2)_{yy}+E(t),
\end{align}
with
\begin{align}
&A(t)=g^2+k^2,\quad
B(\zeta,t)=(\frac{g'}{g}+\frac{k'}{k})\zeta-2k\frac{m}{g}=B_1(t)\zeta+B_0(t),\\
&C(t)=\frac{g'}{2g}+\frac{k'}{2k},\quad
D(t)=-\frac{m^2}{g^2},\quad
E(t)=\frac{g''}{2g}+\frac{k''}{2k}.
\end{align}
To eliminate $B,C,D$ and $E$, we apply the transformation
\begin{align}
\Psi(\zeta,y,t)&=R(t)\exp\big\{i[\theta_2(t)\zeta^2+\theta_1(t)\zeta+\theta_0(t)]\big\}\tilde\Psi(\chi,y,T),\\
\hat W (\zeta,y,t)&=L(t)\tilde W (\chi,y,T)+P(\zeta,t),\\
\chi(\zeta,t)&=\alpha(t)\zeta+\beta(t), \\
\qquad T &=T(t).
\end{align}
$E(t)$ is eliminated if $P$ has the form $\displaystyle P(\zeta,t)=\frac{E(t)}{2A(t)}\,\zeta^2+P_1(t)\zeta+P_0(t)$. The transformation functions to be determined are $\alpha$, $\beta$, $T$, $R$, $\theta_0$, $\theta_1$, $\theta_2$, $L$, $P_0$, $P_1$. We see that  the coefficients $B,C,D$ are eliminated if the following conditions are satified:
\begin{subequations}\label{sixc}
\begin{eqnarray}
   \label{sixc1} &&\dot\theta_2+2B_1\theta_2+4A\theta_2^2+\frac{E}{2A}=0,\\
   \label{sixc2} &&\dot\theta_1+(B_1+4A\theta_2)\theta_1+P_1+2B_0\theta_2=0,\\
   \label{sixc3} &&\dot\theta_0=D-P_0-B_0\theta_1-A\theta_1^2,\\
   \label{sixc4} &&\frac{\dot R}{R}+C+2A\theta_2=0,\\
   \label{sixc5} &&\frac{\dot \alpha}{\alpha}+B_1+4A\theta_2=0,\\
   \label{sixc6} &&\frac{\dot \beta}{\alpha}+B_0+2A\theta_1=0.
\end{eqnarray}
\end{subequations}
See that once it is possible to solve $\theta_2(t)$ from the first equation above which is Riccati-type, the other five determine $\theta_1$, $\theta_0$, $R$, $\alpha$ and $\beta$. This way we arrive at
\begin{subequations}\label{}
\begin{eqnarray}
   \label{} &&i\tilde\Psi_T+\frac{\alpha^2A}{\dot T}\tilde\Psi_{\chi\chi}+\frac{a_1}{\dot T}\tilde\Psi_{yy}=\frac{a_2R^2}{\dot T}|\tilde\Psi|^2\tilde\Psi+\frac{L}{\dot T}\tilde W \tilde \Psi,\\
   \label{} &&\tilde W_{\chi\chi}+\frac{b_1}{\alpha^2A}\tilde W _{yy}=\frac{b_2R^2}{\alpha^2 AL} (|\tilde\Psi|^2)_{yy}.
   \end{eqnarray}
\end{subequations}
We have still the option to choose $L(t)$ and $T(t)$, and the functions $P_0$, $P_1$ are arbitrary. Let us choose  $\dot T=L=\alpha^2A$. That gives us the simplified system
\begin{subequations}\label{21reduced}
\begin{eqnarray}
   \label{} &&i\tilde\Psi_T+\tilde\Psi_{\chi\chi}+\frac{a_1}{\alpha^2A}\tilde\Psi_{yy}=\frac{a_2R^2}{\alpha^2A}|\tilde\Psi|^2\tilde\Psi+\tilde W \tilde \Psi,\\
   \label{} &&\tilde W_{\chi\chi}+\frac{b_1}{\alpha^2A}\tilde W _{yy}=\frac{b_2R^2}{(\alpha^2 A)^2} (|\tilde\Psi|^2)_{yy}.
   \end{eqnarray}
\end{subequations}
Therefore, we have successfully reduced the (3+1)-DS system to a (2+1)-DS system with variable coefficents.

We studied the KMV-algebra of a variable coefficient DS system in \cite{gungor2016variable} and there we include the conditions for a system of  the form \eqref{21reduced} to be transformable to the integrable constant coefficient equation. Let us recall the findings there. \emph{The variable coefficient system }
\begin{subequations}\label{VDS}
\begin{eqnarray}
   \label{DDSa}  &&i\psi_t+p_1(t)\psi_{xx}+p_2(t)\psi_{yy}=q_1(t)|\psi|^2\psi+q_2(t)\psi w, \\
   \label{DDSb}  &&w_{xx}+r_1(t)w_{yy}=r_2(t)(|\psi|^2)_{yy}
\end{eqnarray}
\end{subequations}
\emph{admits the (centerless) Virasoro
algebra as a symmetry algebra if and only if the coefficients satisfy the relations in Table 1.}
\begin{table}\label{pqr}
\caption {Classification results in the integrable case for \eqref{VDS}} \label{tab2}
\begin{center}
    \begin{tabular}{| l | l | l | }
    \hline
    $ p_1(t)=\text{free}$   & $\displaystyle q_1(t)=\frac{p_1(t)}{k_0\int p_1(t)dt+k_1}$   & $\displaystyle r_1(t)=\frac{r_{10}}{(k_0\int p_1(t)dt+k_1)^2}$ \\ \hline
    $\displaystyle p_2(t)=-r_{10}\frac{p_1(t)}{(k_0\int p_1(t)dt+k_1)^2}$ & $q_2(t)=\text{free}$ & $\displaystyle r_2(t)=\frac{r_{20}}{(k_0\int p_1(t)dt+k_1)^3} \frac{p_1(t)}{q_2(t)}$  \\ \hline
        \end{tabular}
\end{center}
\end{table}
Furthermore, \emph{under the conditions of Table 1, the symmetry algebra of the  system \eqref{VDS} is
isomorphic to that of the integrable standard DS system and a point transformation  exists transforming between each other} \cite{gungor2016variable}.

When we check those conditions, we find that transformation of \eqref{21reduced} to the constant coefficient integrable equation requires that \emph{(i)} $b_1=-a_1$ and \emph{(ii)}   the above equations \eqref{sixc} can be solved consistently  with
\begin{equation}
R(t)=(k_0t+k_1)^{1/2}
\end{equation}
where $k_0,k_1$ are arbitrary constants.

\section{Conclusion}
In this work we considered a (3+1)-dimensional Davey-Stewartson type system that was derived in the context of plasma physics. There are slightly different types of equations available in the literature, and the one we chose to work on is the closest to the well-known DS-system in (2+1)-dimensions. We first identified the symmetry algebra of the system to be an infinite-dimensional Lie algebra of Kac-Moody type. Comparing with the (2+1)-dimensional case, the additional terms in the $z$-variable in \eqref{3DDS} causes the Virasoro part of the symmetry algebra to degenerate to  translations and scaling. The Kac-Moody part of the algebra is preserved, by addition of an infinite-dimensional generator for the additional variable $z$. We also have a rotational symmetry in the equation.

Using the infinite dimensional symmetry algebra, we reduced the (3+1)-dimensional system \eqref{3DDS} to the  variable coefficent DS system \eqref{21reduced} which is in (2+1)-dimensions. That has been the main accomplishment of this article. We expect the reductions we have obtained will draw attentions of researchers that search for analytical or numerical solutions to the multidimensional PDE \eqref{3DDS}. Last but not the least, we believe that bringing into attention the  PDE we have considered will serve as a stimulation for further analysis, also for other similar models we have mentioned here, through different approaches.
\section*{Acknowledgement}
I would like to thank Prof. Faruk G\"{u}ng\"{o}r on reading the manuscript and for sharing his valuable comments.

\small

\bibliography{litera}
\bibliographystyle{unsrt}
\bibliographystyle{plain}

\end{document}